\documentclass[conference, 10pt]{IEEEtran}
\IEEEoverridecommandlockouts
\usepackage[T1]{fontenc}
\usepackage[utf8]{inputenc}
\usepackage[english]{babel}
\usepackage{balance}
\usepackage{amsmath, mathtools}
\usepackage{amsfonts,amsthm,bm}
\usepackage{amssymb}
\usepackage{graphicx}
\usepackage{standalone}
\usepackage{mathtools}
\usepackage{color, colortbl}
\usepackage{soul}
\usepackage[acronym,shortcuts]{glossaries}
\usepackage{subcaption}

\def\BibTeX{{\rm B\kern-.05em{\sc i\kern-.025em b}\kern-.08em
 T\kern-.1667em\lower.7ex\hbox{E}\kern-.125emX}}

\newacronym{mdp}{MDP}{Markov decision process}
\newacronym{iot}{IoT}{internet of things}
\newacronym{MTC}{MTC}{machine-type communications}
\newacronym{URLLC}{URLLC}{ultra reliable low latency communications}
\newacronym{mMTC}{mMTC}{massive machine-type communications}
\newacronym{5G}{5G}{fifth-generations}
\newacronym{gNB}{gNB}{new-generation node base}
\newacronym{MTD}{MTD}{machine-type device}
\newacronym{sALOHA}{S-ALOHA}{slotted ALOHA}
\newacronym{NOMA}{NOMA}{non-orthogonal multiple access}
\newacronym{MMPC}{MMPC}{min-max pairwise correlation}
\newacronym{SALOHA}{S-ALOHA}{slotted ALOHA}
\newacronym{RA}{RA}{random access}
\newacronym{LRI}{LRI}{linear reward-inaction}
\newacronym{PDF}{PDF}{probability density function}
\newacronym{SLRI}{S-LRI}{static-LRI}
\newacronym{DLRI}{D-LRI}{dynamic-LRI}
\newacronym{hmm}{HMM}{hidden Markov model}
\newacronym{m2m}{M2M}{machine-to-machine}
\newacronym{iiot}{IIoT}{industrial internet-of-things}
\newacronym{rl}{RL}{reinforcement learning}

\DeclareMathOperator*{\argmax}{argmax}

\title{Coordinated Random Access for Industrial IoT With Correlated Traffic By Reinforcement-Learning} 
\author{Alberto Rech,  and Stefano Tomasin \\ \small Department of Information Engineering, University of Padova, Italy. 
\\Emails: rechalbert@dei.unipd.it, tomasin@dei.unipd.it}

\begin{document}

\bstctlcite{IEEEexample:BSTcontrol}

\maketitle

\begin{abstract}
We propose a coordinated random access scheme for \ac{iiot} scenarios, with \acp{MTD} generating sporadic correlated traffic. This occurs, e.g., when external events trigger data generation at multiple \acp{MTD} simultaneously. Time is divided into frames, each split into slots and each \ac{MTD} randomly selects one slot for (re)transmission, with \acp{PDF} specific of both the \ac{MTD} and the number of the current retransmission. \acp{PDF} are locally optimized to minimize the probability of packet collision. The optimization problem is modeled as a repeated Markov game with incomplete information, and the \acl{LRI} algorithm is used at each \ac{MTD}, which provably converges to a deterministic (suboptimal) slot assignment. We compare our solution with both the \acl{sALOHA} and the \acl{MMPC} random access schemes, showing that our approach achieves a higher network throughput with moderate traffic intensity. 
\end{abstract}

\glsresetall
\section{Introduction}

\Ac{MTC} are considered as a key emerging application of \ac{5G}-and-beyond cellular networks, and the technology should be updated to support them. The sporadic nature of transmissions by a large number of \acp{MTD} make inefficient the current uplink multiple access scheme, based on resource request and grant. Thus, a \ac{RA} solution is to be preferred. Uncoordinated \ac{RA} \cite{Zanella} has been advocated as effective in dealing with collisions, while entailing a limited communication overhead. Still, the high density of \acp{MTD} in \ac{5G}-and-beyond networks highly increases the chances of collisions in absence of coordination. In particular, in \ac{iiot} scenarios, the uplink traffic generated by \acp{MTD} may be highly correlated, as a result of common underlying traffic generation phenomena. For example, close-by temperature sensors in a production line may send signals almost simultaneously, as they sense the same variation of temperature. On one hand, this correlation further increases the chances of collisions, while, on the other hand, it can be exploited to indirectly coordinate \ac{RA}, to satisfy the strict throughput and latency requirements of \ac{iiot} applications.

In the literature, several coordinated \ac{RA} approaches have been proposed. A first and widely used solution is the \ac{sALOHA}, where time is organized in slots, \acp{MTD} transmit at the beginning of the first slot after the packet generation, and, in case of collisions, a random delay is added before retransmission. Typically, the random delay has the same statistics for all \acp{MTD} and the coordination is limited to the synchronization of slots.
An \ac{URLLC} scenario, wherein a set of devices are competing for a limited number of slots in uplink, is considered in \cite{learn2MAC}: an iterative online learning algorithm running at each device updates the slot selection, based on the achieved latent throughput. However, the correlation in the packet generation process is not exploited. Instead, traffic correlation has been considered in an \ac{m2m} scenario \cite{cs_corr}, where \acp{MTD} are clustered and a compressed sensing algorithm is applied to allocate resources to the clusters. Still, clustering entails a significant overhead, greatly reducing the efficiency of the \ac{RA} scheme.
An extreme case of coordinated \ac{RA} is the fast uplink grant, where each \ac{MTD} is assigned a single slot, shared with other \acp{MTD}, thus collisions may still occur. Under a correlated traffic scenario, the \ac{MMPC} scheme \cite{Popovski} assign  slots  by grouping \acp{MTD}  according to their correlation in packet generation. \ac{MMPC} is designed for a system without retransmissions in case of collisions, which is however a useful feature in many scenarios. A traffic prediction-based approach for fast uplink grant is proposed in \cite{Popovski2}, where the packet generation and transmissions  is modeled by an \ac{hmm}, and the slot allocation aims at minimizing the average packet age of information. Also in this case,  retransmissions are not considered. Moreover,  both \cite{Popovski} and \cite{Popovski2} are centralized solutions, where the \ac{gNB} allocates slots and communicates the allocation to \acp{MTD}, thus suffering from a communication overhead. 

In this paper, within a context of cellular system supporting \ac{iiot},  we propose a novel coordinated uplink \ac{RA} solution under correlated traffic: our solution aims at  overcoming the limitations of current coordinated \ac{RA} solutions. Time is divided into frames, each split into slots, and each \ac{MTD} randomly selects a slot for its transmissions. The \ac{PDF} for random slot selection is designed specifically for each \ac{MTD} and for the number of the current retransmission. The \acp{PDF} are obtained by an iterative approach, carried out locally at each \ac{MTD},  to minimize the probability of collision. To this end, we first model the distributed optimization problem as a repeated Markov game with incomplete information, where \acp{MTD} are the players and transmission slots are the actions. Then, we resort to the \ac{LRI} algorithm for the \ac{PDF} optimization. The \ac{LRI} provably converges to a (suboptimal) pure strategy, thus \acp{MTD} will deterministically select the transmission slot, still in different positions for each retransmissions. Lastly, we compare our solution with the \ac{sALOHA} and \ac{MMPC} \ac{RA} schemes, showing that our approach achieves the highest network throughput with moderate traffic correlation and intensity. 

The rest of the paper is organized as follows. In Section~\ref{systemmodel} we introduce the system model of correlated packet generation and slotted coordinated \ac{RA}. The Markov game model describing our distributed optimization problem and the proposed reinforcement-learning algorithm are both presented in Section~\ref{Mathematicalmodeling}. In Section~\ref{numericalresults} we discuss the numerical results and  compare our  \ac{LRI} scheme with existing \ac{RA} schemes. Finally, in Section~\ref{conclusions} we draw some conclusions.
 
{\it Notation:} vectors are denoted in lower-case bold, matrices as uppercase bold. $\mathbb{P}(\cdot)$ and $\mathbb{E}[\cdot]$ are the probability and expectation operators, respectively.

\section{System Model}\label{systemmodel}

We consider a cellular network with $N$ static \acp{MTD}. Each \ac{MTD} transmits in uplink to a \acrfull{gNB}. Time is split into frames, each split into $K$ slots.   We first describe the packet generation procedure, and then the \ac{RA} protocol.

\subsection{Packet Generation} 

Let $y_n(t)$ be the indicator function of packet generation, i.e., $y_n(t) = 1$ if \ac{MTD} $n$ generates a packet at frame $t$, while $y_n(t) = 0$ otherwise; let us also define the row vector $\bm{y}(t) = [y_1(t), \dots, y_N(t)]$. Packet generations are triggered by events common to multiple (random) \acp{MTD}, therefore variables $y_n(t)$, $n \in \{1, \dots, N\}$ are correlated.  In particular, the packet generation statistics is described through the joint probability distribution
\begin{equation}
    \phi(\bm{y}(t)) = \mathbb{P}[y_1(t) = b_1, y_2(t) = b_2, \ldots, y_N(t) = b_N],
\end{equation}
where $b_n \in \{0,1\}$.
Moreover, let $w_n = \mathbb{P}(y_n(t)=1)$ be the marginal probability of packet generation at \ac{MTD} $n$. In Section~IV, we will consider a specific correlated traffic generation model, while the derivation of our proposed \ac{RA} scheme holds in  general for any correlated traffic.

We assume that each \ac{MTD} can store only one packet for transmission and an \ac{MTD} that already stores a packet will drop other generated packets. Packets are generated at the end of each frame and stored (one per \ac{MTD}), then their transmission starts in the next frame. At each frame $t$, packets may be generated at  \acp{MTD} with a joint probability, according to the underlying process (e.g., detection of temperature variation in an industrial line). 

\subsection{\ac{RA} Scheme}

According to a coordinated \ac{RA} protocol, each \ac{MTD} with a stored packet attempts to transmit it in each frame, selecting slot $a_n(t) \in \{1, \dots, K\}$, until either the packet is successfully delivered to the \ac{gNB}, or a maximum number of transmissions $\beta$ is achieved. In this latter case, the packet is discarded. 

Let $x_n(t) = i$, $i>0$, indicate that \ac{MTD} $n$ in frame $t$ is performing the $i$-th transmission attempt of its packet. We also set $x_n(t) = 0$ if $y_n(t) = 0$, and define the vector $\bm{x}(t) = [x_1(t), \dots, x_N(t)]$. If the maximum number of transmission attempts is reached in frame $t$, the packet is discarded, and at the next frame we have $x_n(t+1) = y_n(t)$. 

The probability of \ac{MTD} $n$ transmitting in slot $k$ at the $i$-th attempt is
\begin{equation}
p_{n,k}(i) = {\mathbb P}(a_n(t) = k|x_n(t) = i). 
\end{equation}
Thus, the \ac{PDF} of the slot selected for transmission  by \ac{MTD} $n$ in frame $t$ is $\bm{p}_n(x_n(t))$, which depends of the number of transmissions $x_n(t)$ of the current packet.  

The \acp{PDF} $\bm{p}_n(x_n(t))$ define the \ac{RA} scheme. For example, a \ac{SALOHA} protocol selects the slot uniformly at random and independently for \acp{MTD}, i.e., $p_{n, k}(i)=\frac{1}{K}$. The design of \ac{MTD} \acp{PDF} is the subject of this paper, and will be discussed in the next sections.

\subsection{Collision Model and \ac{gNB} Feedback}

A collision occurs whenever two or more \acp{MTD} schedule their transmissions in the same slot. In this case, we assume that the \ac{gNB} observes an erasure and cannot decode any packet, thus all transmissions fail. In absence of collisions, we assume that the \ac{gNB} always correctly receives the packet. 

Let $z_n(t)$ be the binary variable representing the outcome of the transmission of \ac{MTD} $n$ at frame $t$, thus $z_n(t) = 1$ if the transmission is successful, and $z_n(t) = 0$ otherwise. The success probability at frame $t$ is
\begin{equation}
\begin{split}
q_{n,k}(t) &=\mathbb{P}\big(z_n(t) = 1|a_n(t) = k \big)\\
&=\prod^N_{\substack{\tiny\begin{array}{cc} m = 1 \\ m\neq n \end{array}
 }}\big[1-\mathbb{P}(x_m(t)>0)p_{m,k}(x_m(t))\big].
 \end{split}
\end{equation}
At the end of frame $t$, the \ac{gNB} sends in unicast the acknowledgement to each \ac{MTD} $n$ for which the packet was successfully received.

\paragraph*{Knowledge Assumptions}
The statistics of packet generation are not known and the acknowledgments are sent in unicast, thus the outcome of transmissions is known only to the transmitting \ac{MTD}.

\section{\acp{PDF} Optimization}\label{Mathematicalmodeling}

We now propose a fully distributed algorithm for the optimization of the \ac{MTD}  \acp{PDF}  $\bm{p}_{n}(i), \; i \in \{1, \dots,\beta\}$. The algorithms operates locally at each \ac{MTD}, with the aim of maximizing the \ac{MTD} individual throughput, i.e., minimizing the number of retransmissions.

To this end, we first model the \ac{RA} scheme as a Markov game, wherein \acp{MTD} are the players competing in the slot selection.
Markov games are of particular interest as they represent a specific framework for multi-agent \ac{rl}. Indeed, differently from a \ac{mdp}, wherein a single adaptive agent interacts with the environment and  secondary agents can only be part of it, Markov games allow to model multiple adaptive agents (players) interacting each other for cooperative or competing goals \cite{littman1994markov}.
Several multi-agent \ac{rl} algorithms have been developed to learn equilibrium points in Markov games. For the specific task of slot selection, we resort to \ac{LRI}, a learning automata algorithm, which  learns an equilibrium point of the game by updating the \acp{PDF} at each retransmission. \ac{LRI} provably converges to a sub-optimal deterministic  solutions, leading each \ac{MTD} to always transmit in the same slot when facing a certain transmission attempt. 

\subsection{Slot Selection As a Markov Game}

\paragraph*{Game Definition} Our slot selection process can be modeled as a Markov game (also called stochastic game), where \acp{MTD} are the players, and their actions are their slots selected for transmission. The game is played in multiple {\em rounds}, once per frame. 

The action taken by each \ac{MTD} $n$ in frame $t$ depends only on the number of retransmissions $x_n(t)$, which represents the {\em state} of the player. There are $\beta+1$ states, denoted as $0, 1, \ldots, \beta$, where state 0 indicates that the \ac{MTD} has no packet to transmit, while at state $\beta$  the maximum number of retransmissions is reached. The {\em strategy} of player $n$ is the set of \acp{PDF} by which actions are taken, i.e., $\{\bm{p}_n(i), i=1, \ldots, \beta\}$; note that at state 0 only one action (no transmission) is accessible.

At the end of each round (frame), player $n$ receives the \textit{reward} $z_n(t)$, which depends on the actions of all the players. The \textit{utility function} is the expected reward, which for each \ac{MTD} $n$ can be written as
\begin{equation}\label{utilityfunc}
u_n(\bm{\pi}) = \mathbb{E}[z_n(t)],
\end{equation}
where we highlighted the dependency of the utility from the strategies. The objective of \ac{MTD} $n$, is to find a strategy matrix $\bm{\pi}^*_n$ which maximizes its own expected reward, i.e.,
\begin{equation}\label{optstrategy}
 \bm{\pi}^*_n = \argmax_{\bm{\pi}_n} u_n(\bm{\pi}).
\end{equation}

This is a game of incomplete information, since players have no knowledge on the other players actions.  Each player selects its own strategy with an individual objective, thus the game is non-cooperative. 

\begin{figure} 
 \centering
 \includegraphics[width=.28\textwidth]{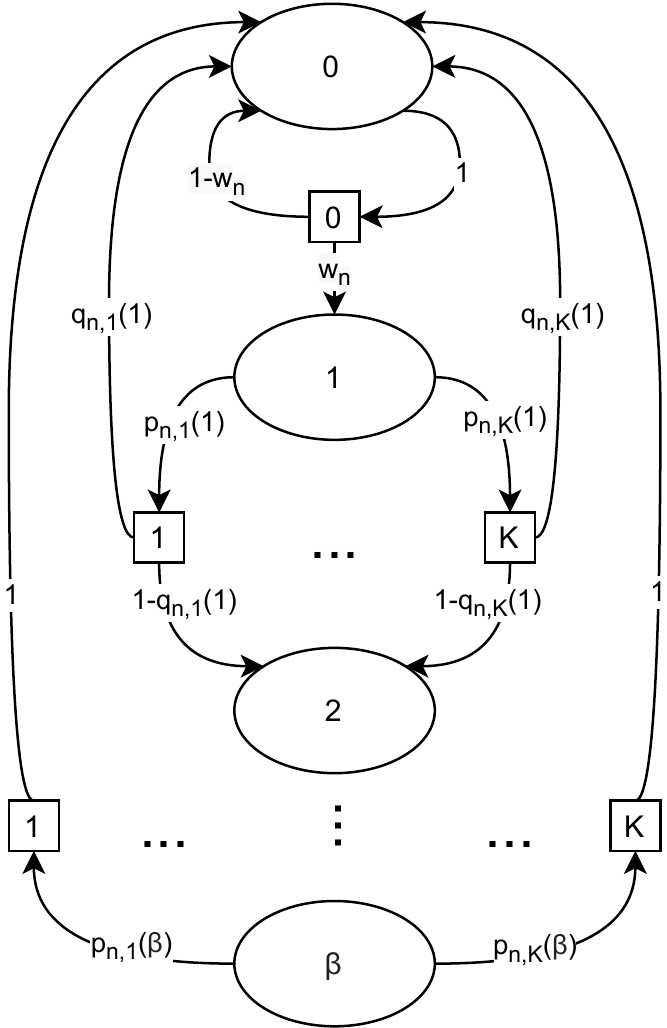}
 \caption{State-action transition diagram of \ac{MTD} $n$. Ellipses denote the states $x_n(t)$, while the squares denote the actions $a_n(t)$.}
 \label{fig:diagram}
\end{figure}

 \paragraph*{State Transitions} We now describe the state transitions, with their probabilities, which are also depicted in Fig.~\ref{fig:diagram}. The transition from state $0$ to state $1$ is due only to a new packet generation, thus occurs with probability $w_n$. States $i = 1, \ldots, \beta-1$, evolve either towards state $0$ (successful transmission) or towards state $i+1$ (failed transmission): the first case occurs with probability $q_{n,k}(i)$ upon action $a_n(t) =k$, while the latter occurs with probability $1 - q_{n,k}(i)$. When in state $\beta$, the packet is either successfully received or discarded and possibly replaced by a new packet: thus, this state evolves with probability $w_n$ to state 1 and with probability $1-w_n$ to state 0. In Fig.~\ref{fig:diagram}, ellipses denote states $x_n(t)$, while  squares denote actions $a_n(t)$, and on the arrows we indicate either the probabilities of taking actions (moving from an ellipse to a square) or the state transition probability for a given action (moving from a square to an ellipse). When in state 0, only one action is possible (no transmission) denoted with 0 in the square.

\paragraph*{State of the Game} The {\em state of the game} at round $t$ is the collection of the states of all players, $\bm{x}(t)$. Let $\bm{\pi}_n$ be the {\em strategy matrix} of player $n$, defined as
\begin{equation}
 \bm{\pi}_n =[\bm{p}_n(1), \dots, \bm{p}_n(\beta)].
\end{equation}
Let us also define the matrix collecting all strategies of each user in each state as  $\bm{\pi} = [\bm{\pi}_1, \dots, \bm{\pi}_N]$.

\subsection{Learning The Strategies}

The objective of each \ac{MTD} is to find a strategy that maximizes the expected reward \eqref{optstrategy} at each transmission attempt. 
To this end, we resort to the \ac{LRI} algorithm \cite{LRI_first}, which is run locally by each \ac{MTD} and works iteratively, one iteration per frame. Let $\{\bm{p}^{(t)}_{n}(i), i=1, \ldots, \beta\}$ be the strategy of \ac{MTD} $n$ at frame $t$, where $\bm{p}^{(t)}_{n}(i) = [p^{(t)}_{n,1}(i), \ldots, p^{(t)}_{n,K}(i)]$.

At the first iteration, we start with a uniform \ac{PDF} for all the \acp{MTD}, i.e.,  $p^{(0)}_{n,k}(i) = \frac{1}{K}$ for all $i\in \{1, \dots, \beta\}$ and $k\in \{1, \dots, K\}$. 

At iteration $t$, \ac{MTD} $n$ (storing a packet) transmits in a random slot selected according to its state and strategy. For failed transmissions ($z_n(t) = 0$), the strategy is not updated, thus $\bm{p}^{(t+1)}_n(i) = \bm{p}^{(t)}_n(i)$, $i=1, \ldots, \beta$. 
 If a packet is successfully received ($z_n(t) = 1$), \ac{MTD} $n$ updates its strategy as follows

\begin{equation}\label{update}
 \begin{split}
     p^{(t+1)}_{n,k}(x_n(t)) = \quad \quad \quad\quad \quad \quad\quad \quad \quad\quad \quad \quad\quad \quad \quad \, \\
     \begin{cases}
       p^{(t)}_{n,k}(x_n(t)) + \alpha z_n(t)[1 - p^{(t)}_{n,k}(x_n(t))] \quad & k = a_n(t),\\
       p^{(t)}_{n,k}(x_n(t)) - \alpha z_n(t) p^{(t)}_{n,k}(x_n(t)) \quad & k \neq a_n(t),
     \end{cases}
  \end{split}
\end{equation}
where $\alpha$ is the \textit{learning rate}, which dictates the speed of the learning process. \acp{PDF} relative to other retransmissions than $x_n(t)$ are left unaltered, i.e., $ \bm{p}^{(t+1)}_n(i) =  \bm{p}^{(t)}_n(i)$, $i \neq x(t)$. From  (\ref{update}) we note that the probability of transmitting in slot $k$ is increased, while the other slots are penalized. 

We remark that \ac{LRI} does not require any knowledge of the other players states and strategies. In fact, from \eqref{update}, it can be seen that the algorithm is fully distributed.

\subsection{\ac{LRI} Convergence}

It is proven that, for small values of $\alpha$, the \ac{LRI} algorithm converges to a pure Nash equilibrium \cite{learningNE}, i.e., the strategy of any player $n$ maximizes its utility function, given the strategies of all other players~\cite{tadelis2013game}. Moreover, \ac{LRI} converges to a {\em pure strategy}, i.e., only one slot is deterministically selected by each \ac{MTD} at each retransmission. 

However, note that \ac{LRI} may not provide to the maximum utility $u_n(\bm{\pi})$, and in general will not even provide the maximum sum of utility among all \acp{MTD}. Still, it converges  to a deterministic policy, ensuring the stability of the algorithm.

\section{Numerical Results}\label{numericalresults}

 \begin{figure} 
 \centering
 \begin{subfigure}{\linewidth}
 \includegraphics[width = .9\linewidth]{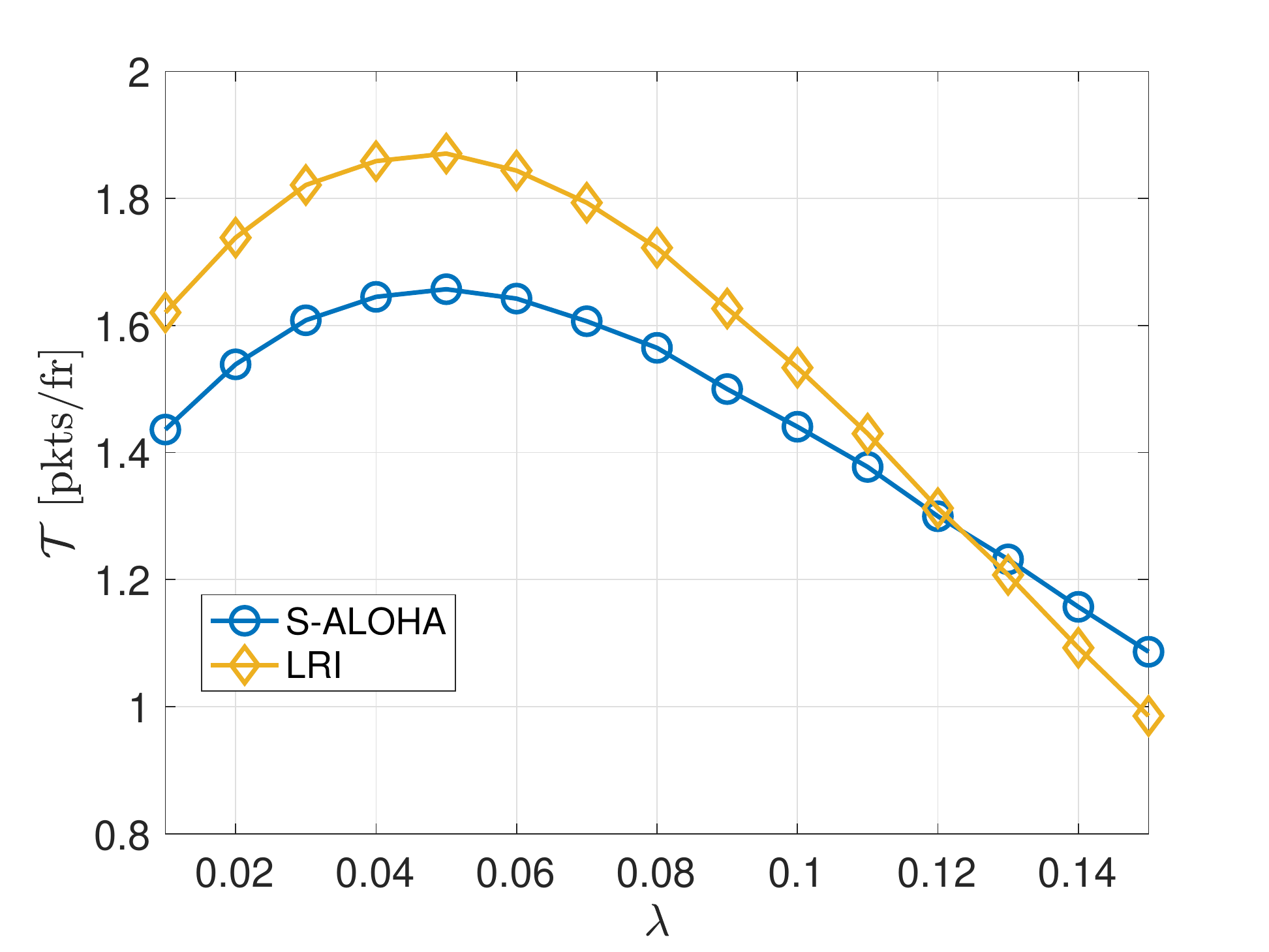}
 \caption{Average system throughput $\mathcal{T}$.}
 \label{fig:DLRImu0_th}
 \end{subfigure}
  \begin{subfigure}{\linewidth}
 \includegraphics[width = .9\linewidth]{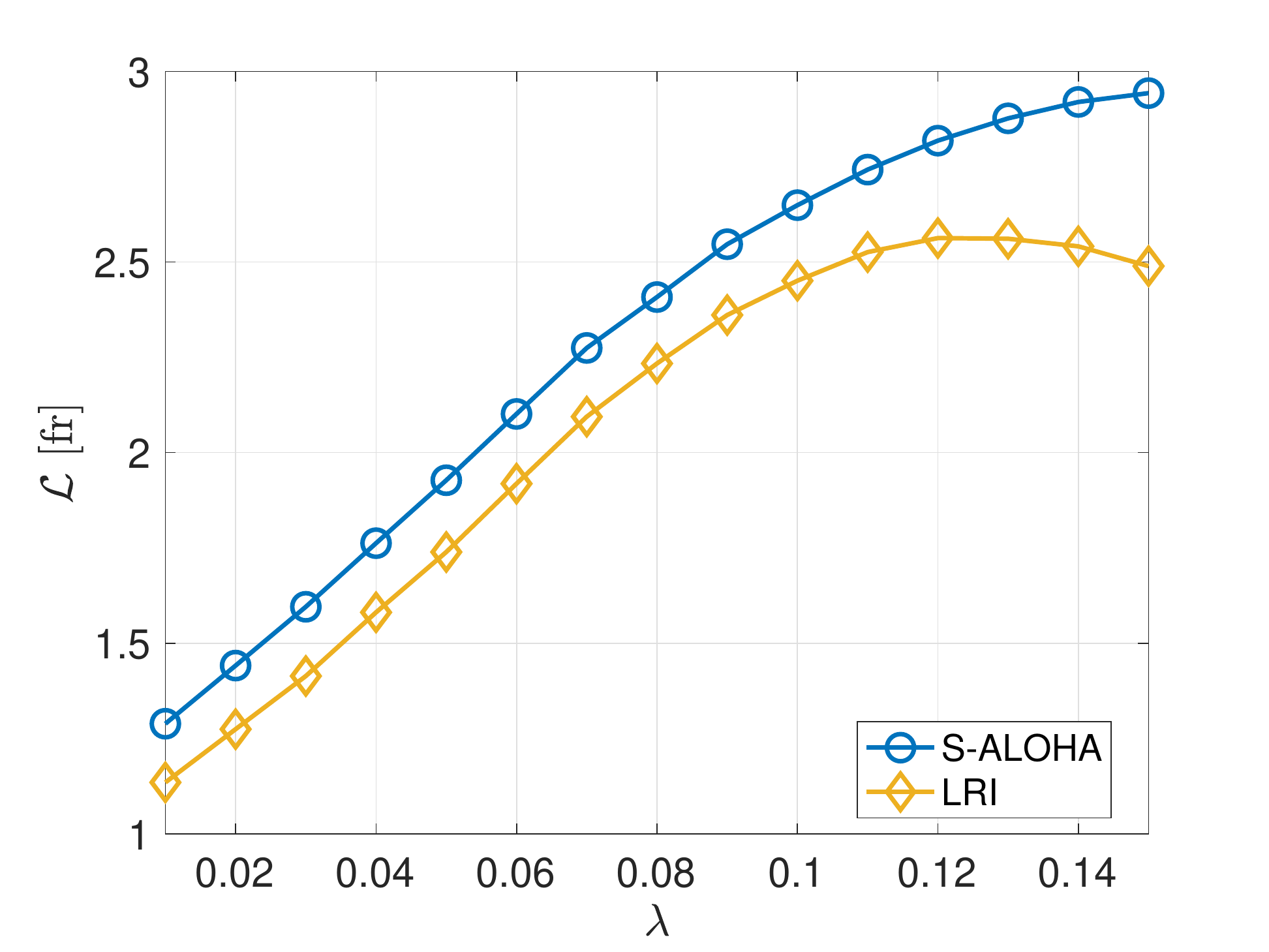}
 \caption{Average packet transmission time $\mathcal{L}$.}
 \label{fig:DLRImu0_L}
 \end{subfigure}
  \caption{Average system throughput (a) and packet transmission time (b) of \ac{LRI} and \ac{SALOHA} as a function of $\lambda$, for $\mu = 0$ and $\beta=5$.}
  \label{fig:DLRImu0_Ltot}
 \end{figure}

\begin{figure} 
 \centering
 \begin{subfigure}{\linewidth}
    \includegraphics[width = .9\linewidth]{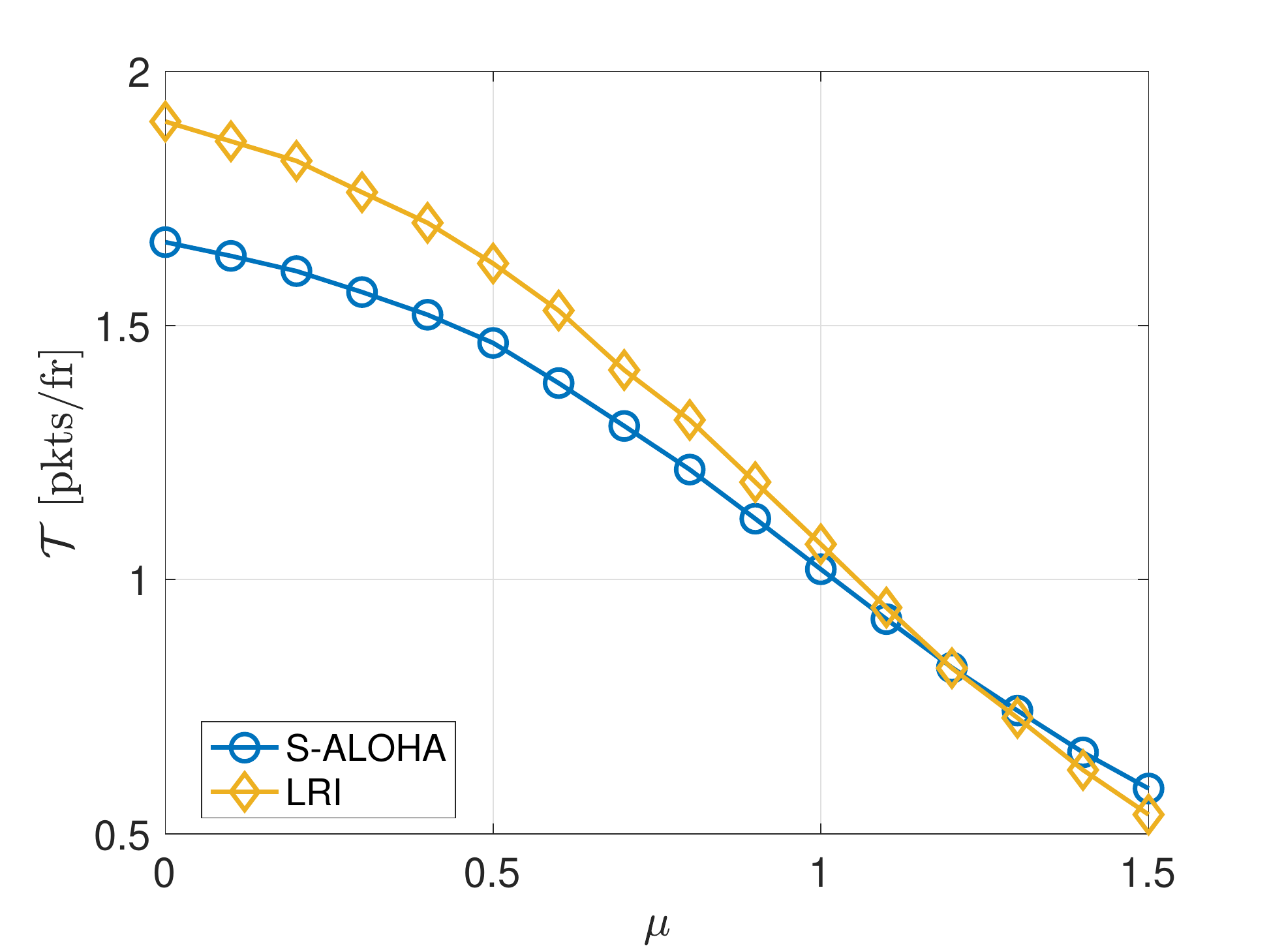}
    \caption{Average system throughput $\mathcal{T}$.}
    \label{fig:DLRImu_th}
 \end{subfigure}
  \begin{subfigure}{\linewidth}
 \includegraphics[width = .9\linewidth]{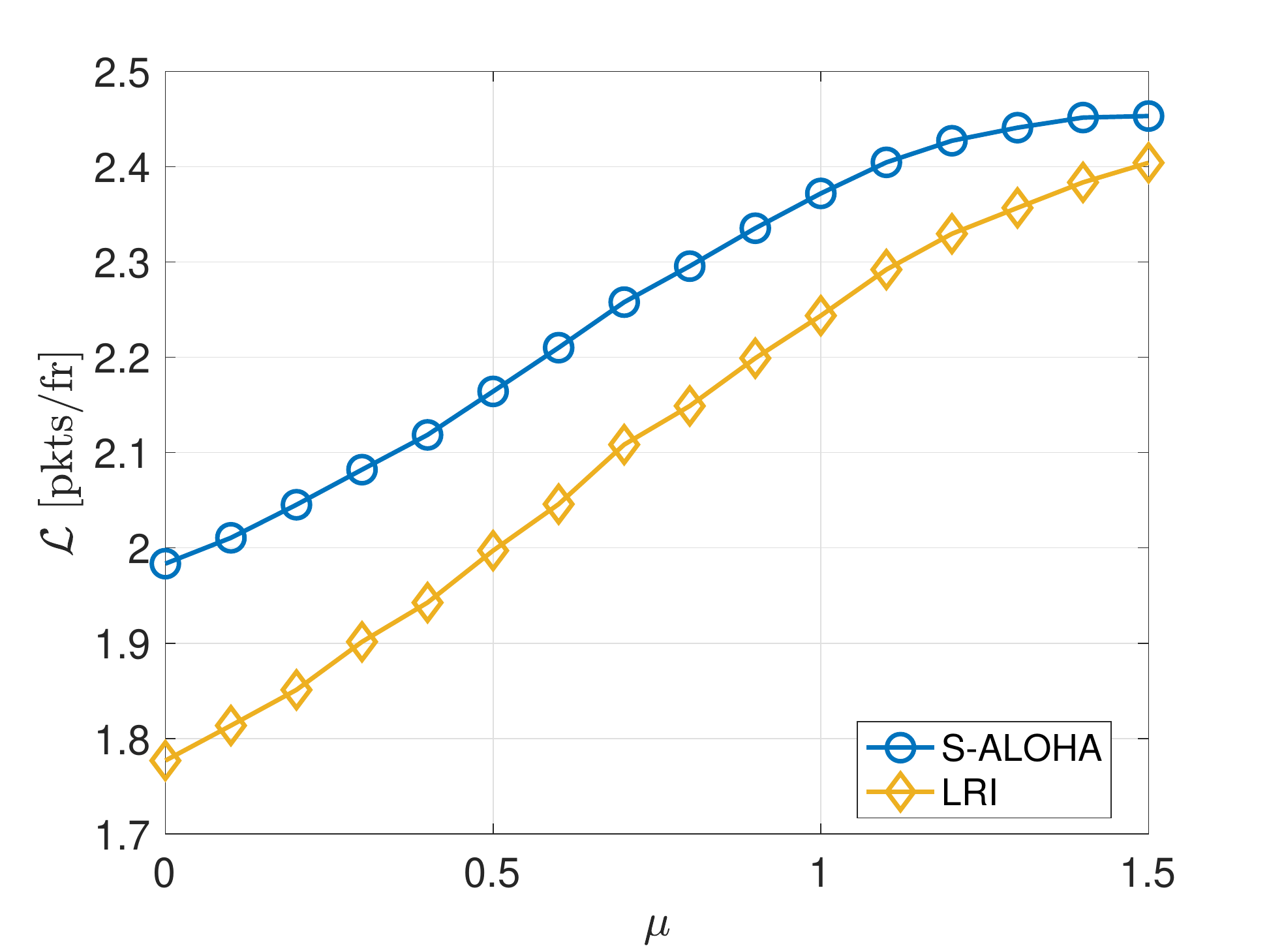}
 \caption{Average packet transmission time $\mathcal{L}$.}
 \label{fig:DLRImu_L}
 \end{subfigure}
  \caption{Average system throughput (a) and packet transmission time (b) of \ac{LRI} and \ac{SALOHA} as a function of $\mu$, for $\lambda = 0.05$ and $\beta = 5$.}
  \label{fig:DLRImu}
 \end{figure}

To assess the performance of the proposed solution, we consider a network where each frame comprises $K = 4$ slots and  $N = 20$ \acp{MTD} are uniformly randomly placed in square area of side $10$~m. 

We consider the space-time Poisson process traffic model of \cite{Popovski}, modified here to take into account packet generations at multiple frames. First, active frames, when packets are generated, are modeled by a temporal Poisson process of intensity $\mu$. In an active frame, several positions ({\em events}) are selected in the area, according to a space Poisson point process of intensity $\lambda$, and all \acp{MTD} within $1.25$~m from an event generate one packet. Note that as $\lambda$ increases, we have two effects: the increase of correlation in packet generation and a higher average number of generated packets in the cell. 
 
Performance is assessed in terms of average packet transmission time (delay) and system throughput. 
In formulas, the average packet transmission time (in frames) is 
\begin{equation}\label{txtime}
 \mathcal{L} = \frac{1}{N}\sum_{n = 1}^{N}\mathbb{E}[x_n(t)|z_n(t)=1],
\end{equation}
while the average system throughput is defined as the ratio between the average number of packets successfully received at the \ac{gNB} and the average number of frames used for its transmission, i.e.
\begin{equation}\label{throughput}
 \mathcal{T} = \frac{1}{\mathcal{L}} \sum_{n=1}^{N}{\mathbb{E}\Big[z_n(t)\Big]}.
\end{equation}
Note that, in our \ac{LRI} scheme, each \acp{MTD} acts in a {\em selfish} fashion, thus an optimal result for the global system throughput \eqref{throughput} cannot be obtained in general.

\subsection{Low-Traffic Scenario}

We first consider a low-traffic scenario ($\mu =0$), where packets are generated only in frame $t=0$. The maximum number of transmission attempts is here $\beta = 5$. 
 
Fig.s~\ref{fig:DLRImu0_th} and \ref{fig:DLRImu0_L} show the average system throughput $\mathcal T$ in packets per frame [pkts/fr] and the average packet transmission time $\mathcal L$ in frames [fr], respectively, both as a function of the spatial events generation rate $\lambda$. The performance according to both metrics is reported for both the  \ac{LRI} and \ac{SALOHA} \ac{RA} schemes. We observe that our \ac{LRI} solution outperforms \ac{SALOHA} with low and moderate event generation rates $\lambda$, while the performance decreases for high values of $\lambda$. Indeed, going from small to moderate event rates, the traffic correlation increases, a condition exploited by \ac{LRI}, which yields a higher throughput than \ac{SALOHA}. For high values of $\lambda$, instead,  the overall generation rate increases (more packets are generated), which increases collisions and ultimately decreases the rate and increases the packet delay, as well known for these kinds of \ac{RA} schemes. Moreover, notice that, for very high values of $\lambda$, the average transmission time is reduced, due to the increase of the probability of packet expiration (maximum number of transmission attempts reached).

Indeed, as \ac{LRI} in general does not find the optimal maximum throughput solution, in this case it turns out to be suboptimal also with respect to \ac{SALOHA}. 

\subsection{Throughput vs Traffic Intensity}
 
We then consider various traffic intensity scenarios, and Fig.~\ref{fig:DLRImu_th} and Fig.~\ref{fig:DLRImu_L} show the performance as a function of $\mu$, for $\lambda = 0.05$.

Note that new packets are generated on average every $1/\mu$ frames, and we always generate packets at frame $t=0$. 
 
In this scenario, whenever a packet is generated at \ac{MTD} $n$ while another is in its buffer, the old packet is dropped and the counter of transmission attempts restart. We note that as $\mu$ increases, the average delay increases and the throughput decreases: this is due to the fact that more  packets yield more collisions, with a throughput reduction. Moreover, we observe that the gain of \ac{LRI} over \ac{sALOHA}, in terms of both delay and throughput, vanishes as $\mu$ increases. Taking for example the first frames, with a higher $\mu$, there will be new packet generations at frame $t>0$, when some \acp{MTD} are still handling packets generated at $t=0$, therefore the statistics of $\bm{x}(t)$ is altered by the new arrivals, decorrelating the resulting  traffic. Packet drops due to new arrivals have a significant impact also on the average delay: indeed, from Fig.~\ref{fig:DLRImu_L}, we observe that the curves are nearly flat for very high values of $\mu$. In this case, \ac{LRI} loose its advantage over \ac{sALOHA}.

\begin{figure}
 \centering
 \includegraphics[width = .9\linewidth]{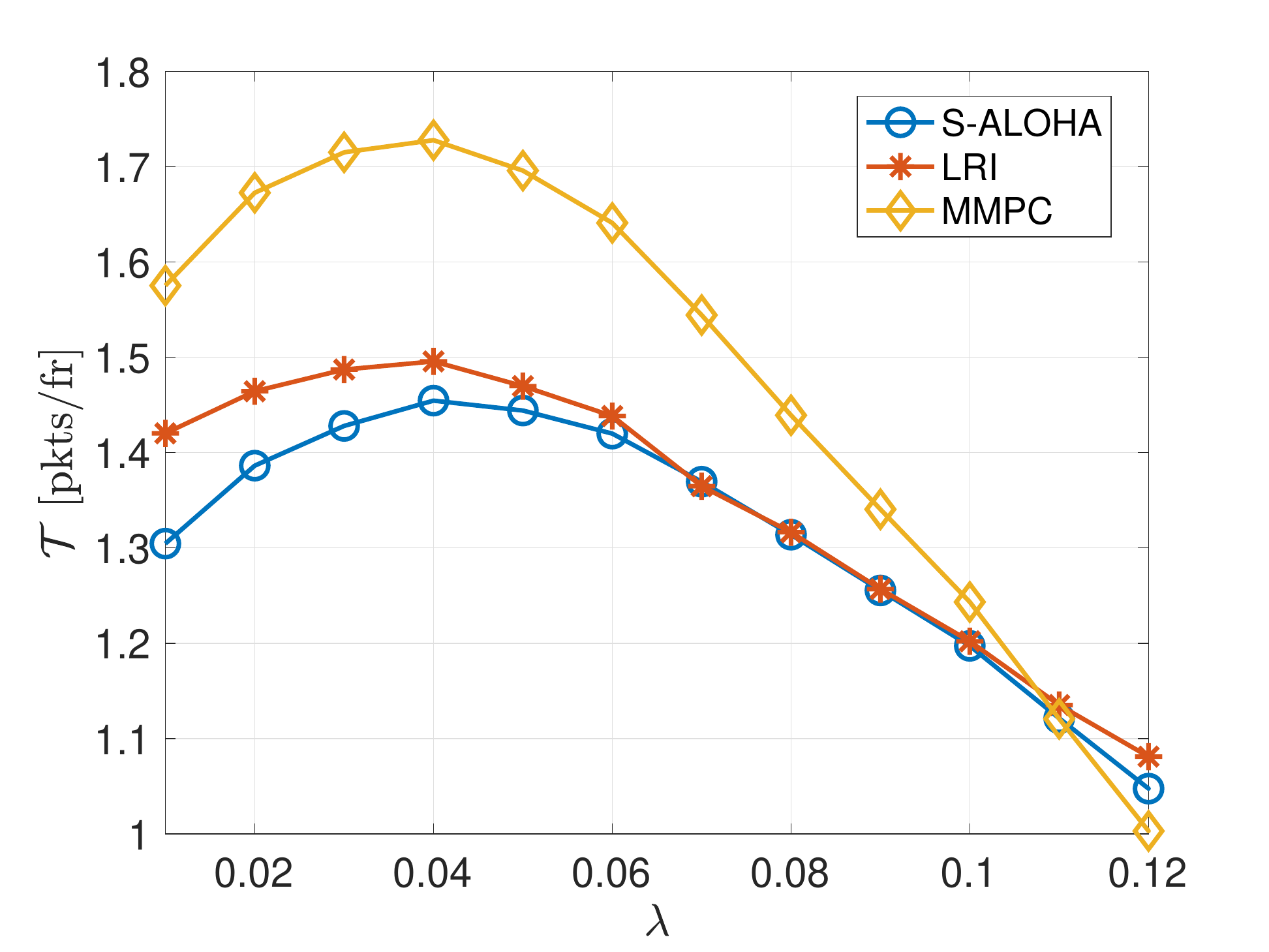}
 \caption{Average system throughput of \ac{LRI}, \ac{MMPC}, and \ac{SALOHA}, as a function of $\lambda$ for $\beta = 1$.}
 \label{fig:SLRI}
 \end{figure}

\subsection{Single Transmission}

We now consider the case $\beta = 1$, which provides a direct comparison with the \ac{MMPC} scheme of \cite{Popovski}: in this case, all colliding packets are discarded without further retransmissions. From \eqref{txtime}, we have $\mathcal{L} = 1$, therefore the system throughput $\mathcal{T}$ boils down to the expected number of successful transmissions in a frame. Fig.~\ref{fig:SLRI} shows the average system throughput of our \ac{LRI}, \ac{MMPC}, and \ac{SALOHA}, as a function of the event generation rate $\lambda$, for $\beta = 1$. The throughput behaviour is similar to that with $\beta > 1$, providing a higher improvement for low event generation rates, while being overcome by \ac{MMPC} and \ac{SALOHA} for higher event generation rates. Indeed, we note that, although both \ac{LRI} and \ac{MMPC} are designed taking into account the traffic correlation, \ac{LRI} has better performance up to moderate event generation rates, as it changes the \ac{MTD} strategies at each retransmission. Again, we observe that for high values of $\lambda$ all \ac{RA} schemes achieve a  lower throughput, with \ac{LRI} degrading its performance due to the selection of a suboptimal solution. 
 
 \begin{figure} 
 \centering
 \includegraphics[width =.9\linewidth]{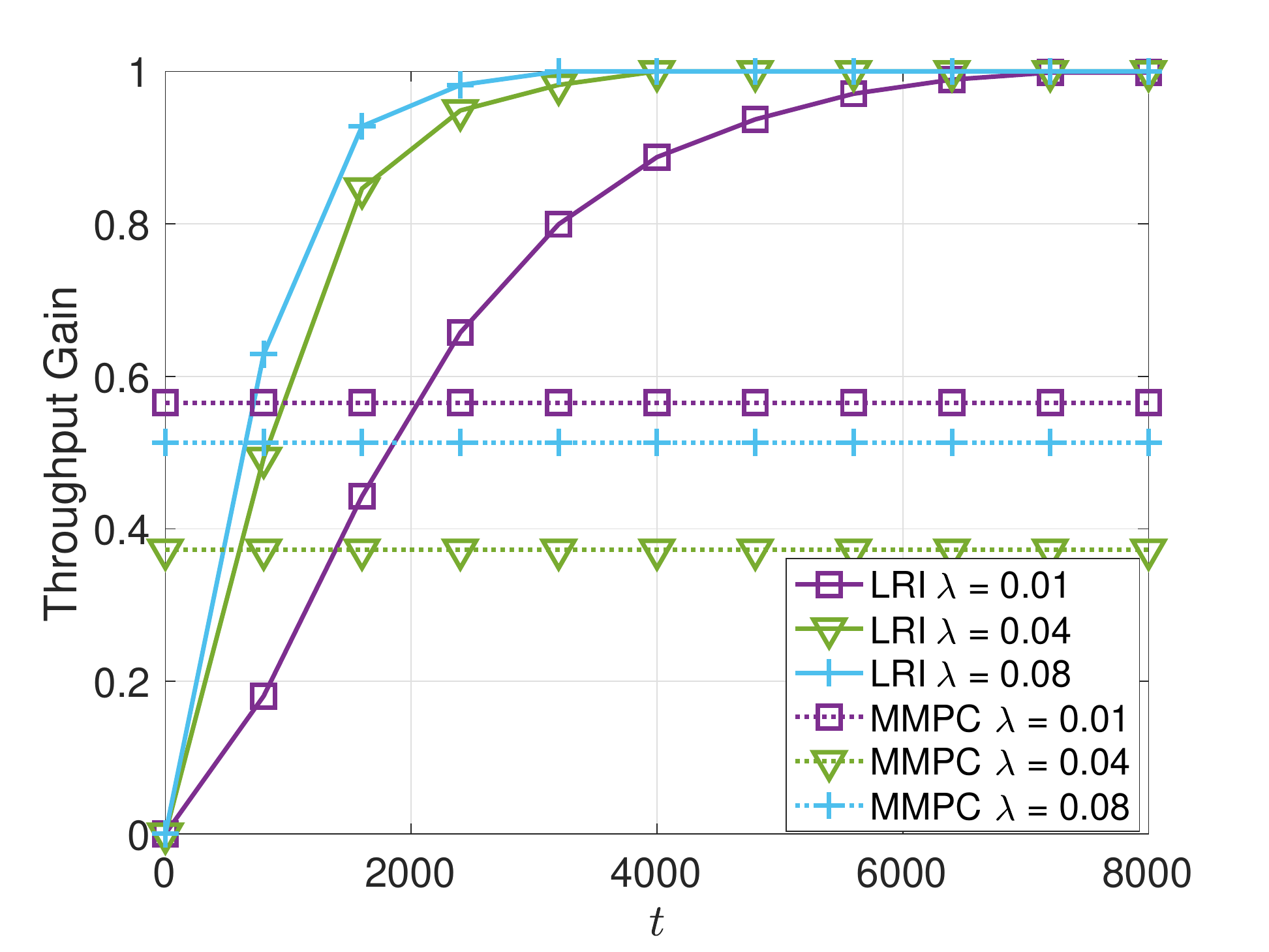}
 \caption{Throughput gain of \ac{LRI} as a function of the frame $t$, for different traffic correlations ($\lambda = 0.01$, 0.04, and 0.08) and $\beta=1$.}
 \label{fig:th_ev}
 \end{figure}
\balance

\subsection{Convergence Speed for Single Transmission}

Finally, we evaluate the convergence speed of the learning algorithm with $\beta = 1$, which still allows the comparison with \ac{MMPC}. For the training, we set the learning rate $\alpha = 0.01$. Let us define the {\em throughput gain}  
\begin{equation}\label{thgain}
 \mathcal{G}_\mathcal{T}(t) = \frac{\mathcal{T}(t) - \mathcal{T}_{\rm S-ALOHA}}{\mathcal{T}_{\rm LRI} - \mathcal{T}_{\rm S-ALOHA}},
\end{equation}
where $\mathcal{T}(t)$ is the throughput computed after $t$ frames of learning, $\mathcal{T}_{\rm S-ALOHA}$ is the throughput of \ac{sALOHA} and $\mathcal{T}_{\rm LRI}$ is the throughput of \ac{LRI} at convergence. Note that we initialize the \ac{LRI} algorithm with uniform \acp{PDF}, thus $\mathcal{T}(0)=\mathcal{T}_{\rm S-ALOHA}$ and $\mathcal{G}_\mathcal{T}(0) = 0$. At convergence, we have $\mathcal{G}_\mathcal{T}(t) = 1$. Fig.~\ref{fig:th_ev} shows the throughput gain of \ac{LRI}, as a function of the learning frames. For comparison purposes, we also report the throughput gain (normalized to the \ac{LRI} throughput) of \ac{MMPC}, obtained by replacing $\mathcal{T}(t)$ with the \ac{MMPC} throughput in \eqref{thgain}. Three event generation intensities are considered, $\lambda = 0.01$, 0.04, and 0.08. We observe that convergence is faster for high values of $\lambda$, as the correlation is in this case stronger, thus the \ac{LRI} iterations quickly adjust the strategy. Moreover, \ac{LRI} already outperforms \ac{MMPC} within about 1\,000 frames. The learning process is slower for low values of $\lambda$, requiring more than 2\,000 frames to overcome the throughput achieved with \ac{MMPC}. 
  
\section{Conclusions}\label{conclusions}
In this paper, we derived a coordinated \ac{RA} scheme for an \ac{MTC} scenario with traffic correlation. We modelled each \ac{MTD} as a player of a Markov game of incomplete information and, applying the \ac{LRI} algorithm, we derived pure Nash equilibrium strategies for each player. Numerical results show that our proposed \ac{LRI} solution outperforms the state-of-the-art \ac{RA} schemes for moderate traffic correlation and intensity.

\bibliographystyle{IEEEtran}
\bibliography{bibs/references}

\end{document}